\documentclass[a4paper,10pt]{article}

\usepackage{amssymb}
\usepackage{amsthm}

\newtheorem{definition}{Definition}
\newtheorem{theorem}{Theorem}
\newtheorem{proposition}{Proposition}

\newcommand{\Ddot}{\,\smash{\stackrel{d}{\bullet}}\,}
\newcommand{\Dwedge}{\,\smash{\stackrel{d}{\wedge}}\,}

\begin{document}

\title{Electromagnetic Duality Based on Axiomatic Maxwell Equations}

\author{Juan  M\'endez\\ \\
Edif. Inform\'atica y Matem\'aticas, Univ. Las Palmas\\ 35017 Las Palmas,
Canary Islands, Spain.\\ E-mail:{\tt jmendez@dis.ulpgc.es}}
\date{}

\maketitle

\begin{abstract}
No positive result has been obtained on the magnetic monopoles
search. This allows to consider different theoretical approaches
as the proposed in this paper, developed in the framework of the
Einstein General Relativity. The properties of second rank
skew-symmetrical fields are the basis of electromagnetic theories.
In the space-time the Hodge duality of these fields is narrowly
related with the rotations in the SO(2) group. An axiomatic
approach to a dual electromagnetic theory is presented. The main
result of this paper is that the stress-energy tensor can be
decomposed on two parts: the parallel and the perpendicular. The
parallel part is easily integrated on the Lagrangian approach,
while some problems appears with the perpendicular part. A
solution with the parallel part alone is found, it generates a
non-standard model of magnetic monopoles neutral to the electric
charges.
\end{abstract}

Pacs: 03.50.De, 04.40.Nr, 14.80.Hv

\section{Introduction}

The concept of magnetic monopole extends the meaning of the
Maxwell equations, by providing a symmetric view of electric and
magnetic fields. The asymmetry of these equations has been for a
long time a motivational phenomenon in both the
theoretical\cite{Blagojevic:1988sh, Giacomelli:1998wk} and
experimental\cite{Caso:1998tx} aspects. According with Hawking and
Ross\cite{Hawking:1995ap}, the belief that electrically and
magnetically charged black holes have identical quantum properties
provides a new interest in all questions related with
electromagnetic duality\cite{Lee:1997pn, Lee:1992vy,
Ignatev:1996qz, Deser:1997xu}. Additionally, as asserted by
Lee\cite{Lee:1997pn}, the magnetic monopole dynamic is related to
strongly interacting elementary particles. Most of the works in
electromagnetic duality are dealing with quantum aspect, while as
appointed by Israelit\cite{Israelit:1996cg, Israelit:1996ie},
little attention is considered to construct a correct classical
theory in the General Relativistic arena. The classical theory can
be considered as a limit of quantum theory, therefore advances in
the first can be useful to the second\cite{Chan:1995bp}. The
construction of a classical theory of electromagnetic duality is
not a closed matter. A pending problem deals with the duality
invariance of both the stress-energy tensor and the Lagrange
action\cite{Hawking:1995ap, Deser:1997xu, Deser:1997gq}.

In classical field theory, if the electromagnetic field is
constructed from a potential vector as: $F_{ab} =
2\nabla_{[a}A_{b]}$, the vanishing of the second set of Maxwell
equations, $\nabla_{[a}F_{bc]}=0$, is due to a property of the
Riemann curvature tensor: $R^{d}_{\;[abc]}=0$. The definition of
the field or the affine properties of the space-time must be
changed to obtain a full electromagnetic duality with symmetrical
Maxwell equations. First type of theories can include different
field definitions from the potential vector, as in the Yang-Mills
model based on gauge theories\cite{Lee:1992vy, Gockeler:1989,
Baez:1994}. Second type of theories can include torsion properties
in the space-time, as based on Weil geometry\cite{Israelit:1996cg,
Dirac:1973}. Any theory which modifies some of these aspects can
generate a non-vanishing second set of Maxwell equations, and
consequently some type of magnetic monopolar current.

Electromagnetic duality is related with the symmetrical role of
electric and magnetic fields in the Maxwell equations. The Hodge
duality is an useful tool to deal with this equation symmetry. It
becomes a cornerstone to connect the physical and the mathematical
concepts of duality. The double duality transformation in
four-dimensional space-time generates an identity with negative
sign. From far, this fact has been qualitatively associated with
the rotation in a plane\cite{Misner:1973}(p 108). Recently an
increased interest arises in the symplectic structure of the
electromagnetic equations\cite{Kachkachi:1997qu,
Chruscinski:1999bi} and also in the SO(2) symmetrical group
embedded in the duality\cite{Wotzasek:1998rj, Chruscinski:1999zw,
Deser:1997mz, Igarashi:1998hd}. Chan and
Tsou\cite{Chan:1997ni,Chan:1999bj} show some of the problems with
duality in the classical theory, and extend it to nonabelian gauge
theory.

In the classical domain of General Relativity the electromagnetic
field can be represented by the skew-symmetrical strength tensor
$F_{ab}$, while in a quantum domain it must be potential based due
to the Bohm-Aharonov effect\cite{Baez:1994, Chan:1999bj,
Aharonov:1959fk}. In this paper a non-quantum approach is
presented, the electromagnetic duality is obtained from the formal
properties of the skew-symmetrical tensors rather than induced
from an a priori magnetic monopole model.

The Hodge duality is a cornerstone of electromagnetic duality. On
a space-time with $D=4k$ dimensionality, it generates a $SO(2)$
symmetrical group\cite{Wotzasek:1998rj, Deser:1997mz}. The other
cornerstone is the Poincar\'e Lemma. It defines the necessary and
sufficient condition for the existence of solutions involving
differential forms. In this paper the solution to the general
symmetrical Maxwell equations is fragmented on pieces based on
closed forms. This is achieved using a two potential approach also
used previously in electromagnetic duality\cite{Zwanziger:1971hk,
Mukherjee:1998he}.

The plan of this paper is as follows. The Sec. 2 contains most of the
mathematical tools used along the paper. It is being presented in a formal way
comprising a set of definitions, propositions and theorems, including also most
of the deductive machinery. This section is partially an experiment in order to
find a reduced set of mathematical elements needed to construct the basis of an
axiomatic electromagnetic theory. The Sec. 3 presents the equations of the
electromagnetic field as are deduced from the axiomatic. The Sec. 4 deals with
the duality invariance, including an invariant formulation for the Maxwell
equations and the stress-energy tensor. The Sec. 5 presents a model of Lagrange
function. Finally, The Sec. 6 is related with the particle-field and
particle-particle interactions.

\section{The Axiomatic of Class $\mathcal{EM}$}

The tensor class $\mathcal{EM}$ is described as a tensor family
with a common property. A set of definitions, propositions and
theorems are formally presented, all are related with equations
involving second rank skew-symmetric tensors. This section is
linearly presented, such as most of the propositions and theorem
become immediately proved from previous assertions. As in General
Relativity a $(M,g_{ab})$ model of the space-time is considered,
where $M$ is a four-dimensional $C^{\infty}$ connected manifold
and $g_{ab}$ is a metric on $M$ with signature: $\{-1,1,1,1\}$. A
p-form and a skew-symmetric tensor of rank p are equivalent
representations used along this section as:
\begin{equation}
 \mathbf{X} =
\frac{1}{p!}X_{a_{1} \cdots a_{p}} {\mathbf{w}}^{a_{1}} \wedge \cdots \wedge
{\mathbf{w}}^{a_{p}} \Leftrightarrow X_{[a_{1} \cdots a_{p}]}
\end{equation}

in a general basis, while for a coordinate basis
$\mathbf{w}^{k}=\mathbf{d}x^{k}$. The wedge product $\wedge$ of a
p-form and a q-form is a $(p+q)$-form which verifies:
$\mathbf{X}\wedge\mathbf{Y} =
(-1)^{pq}\mathbf{Y}\wedge\mathbf{X}$. The dual of a p-form, $
\mathbf{X}$, is a $(n-p)$-form, $^{*}\mathbf{X}$, where $n=4$ and
$0 \le p \le n$. It is defined as\cite{Gockeler:1989}(p 36):
\begin{equation}
^{*}\mathbf{X} = \frac{1}{(n-p)!}\left [\frac{1}{p!}\epsilon_{a_{1} \cdots
a_{n}} X^{a_{1} \cdots a_{p}} \right ] \mathbf{w}^{a_{p+1}} \wedge \cdots
\wedge \mathbf{w}^{a_{n}}
\end{equation}

where $\epsilon_{abcd}$ is the Levi-Civita fourth rank completely
skew-symmetric tensor, with $\epsilon_{0123}=\sqrt{-g}$, and $g$ is the metric
determinant. Throughout this paper the dual of first, second and third rank
tensors will be considered, based on the previous expression they are defined
as\cite{Misner:1973}(p 88):
\begin{equation}
 ^{*}A_{bcd}  =  \epsilon_{abcd} A^{a} \qquad
 ^{*}B_{cd}  =  \frac{1}{2}\epsilon_{abcd}B^{ab} \qquad
 ^{*} C_{d}  =  \frac{1}{3!} \epsilon_{abcd}C^{abc}
\end{equation}

The double duality verifies that: $^{**}\mathbf{X} =
(-1)^{p-1}\,\mathbf{X}$. It is verified that:
$^{**}\mathbf{A}=\mathbf{A}$, $^{**}\mathbf{B} = -\mathbf{B}$ and
$^{**}\mathbf{C}=\mathbf{C}$.

\subsection{Linear Expressions}

For an 1-form $\mathbf{U}=U_{a}\mathbf{w}^{a}$ the exterior
derivative can be expressed from the tensorial derivative as:
$(d\wedge\mathbf{U})_{ab}=(d\otimes \mathbf{U})_{ab} - (d\otimes
\mathbf{U})_{ba}$, so if: $d\otimes \mathbf{U}=(\nabla_{a}
U_{b})\mathbf{w}^{a}\otimes \mathbf{w}^{b}$, it is obtained that:
\begin{equation}
d\wedge\mathbf{U}=d\mathbf{U}=(\nabla_{a} U_{b} - \nabla_{b}
U_{a})\mathbf{w}^{a}\wedge \mathbf{w}^{b}
\end{equation}

For the general case, the derivative of a p-form is a $(p+1)$-form
expressed in terms of the covariant derivative:
\begin{equation}\label{skew}
\mathbf{Y}=d\mathbf{X} \Leftrightarrow Y_{a_{1} \cdots a_{p}}=
\nabla_{a_{1}} X_{a_{2} \cdots a_{p}} - \nabla_{a_{2}} X_{a_{1}
\cdots a_{p}} + \cdots (-1)^{p}\nabla_{a_{p}} X_{a_{1} \cdots
a_{p-1}}
\end{equation}

For a scalar $\phi$, an 1-form $\mathbf{U}$ and a 2-form
$\mathbf{X}$, it is verified that:
\begin{eqnarray}
d\phi & \Leftrightarrow & \nabla_{a}\phi \\ d\mathbf{U} & \Leftrightarrow &
\nabla_{a} U_{b} - \nabla_{b} U_{a}= 2 \nabla_{[a} U_{b]}\\ d\mathbf{X} &
\Leftrightarrow & \nabla_{a} X_{bc} - \nabla_{b} X_{ac} + \nabla_{c} X_{ab} =
3\nabla_{[a} X_{bc]}
\end{eqnarray}

It is verified that $dd=0$, that is $d\mathbf{Y}=0$ in the
equation (\ref{skew}). The co-derivative $\delta \mathbf{X}$ of a
p-form is a $(p-1)$-form expressed as: $\delta \mathbf{X} = \,
^{*}d ^{*}\mathbf{X}$ in the four-dimensional
space-time\cite{Gockeler:1989}(p 40). For an 1-form is verified
that:
\begin{equation}
\mathbf{Z}=\delta\mathbf{X} \Leftrightarrow Z =- \nabla^{a} X_{a}
\end{equation}

If $\phi$ is a scalar: $\delta d \phi \Leftrightarrow - \square \phi$, where
$\square = \nabla^{a} \nabla_{a}$ is the second order differential operator.
Also for a 2-form:
\begin{equation}
\mathbf{Z}=\delta\mathbf{X} \Leftrightarrow Z_{a} = - \nabla^{b} X_{ba}
\end{equation}

If $\mathbf{U}$ is an 1-form in a curved space-time, it is verified that:
\begin{equation}
\delta d \mathbf{U} + d \delta \mathbf{U} \Leftrightarrow - \square U_{a} +
R^{b}_{a} U_{b}
\end{equation}

where $R^{b}_{a}$ is the Ricci tensor. It is also verified that: $\delta\delta
= 0$, that is $\delta \mathbf{Z}=0$ in the previous cases.

\begin{definition}
A second rank tensor $\mathbf{X} \in \mathcal{EM}_{0}$ if it is
skew-symmetric, it means that: $X_{ab} + X_{ba}=0$.
\end{definition}

\begin{definition}
A tensor $\mathbf{X} \in \mathcal{EM}_{1}$, also called an
$\alpha$-field, if $\mathbf{X} \in \mathcal{EM}_{0}$ and also it
verifies: $d\mathbf{X}=0$, equivalent to: $\nabla_{[a}X_{bc]} =
0$.
\end{definition}

\begin{definition}
A tensor $\mathbf{X} \in \mathcal{EM}_{2}$, also called a
$\beta$-field, if $\mathbf{X} \in \mathcal{EM}_{0}$ and also it
verifies: $\delta \mathbf{X}=0$, equivalent to: $ \nabla_{b}X^{ab}
= 0$.
\end{definition}

These classes verify that: $\mathcal{EM}_{1} \subset \mathcal{EM}_{0}$ and
$\mathcal{EM}_{2} \subset \mathcal{EM}_{0}$. Two new classes can be introduced,
the first as intersection of the previous, $\mathcal{EM}_{3} = \mathcal{EM}_{1}
\cap \mathcal{EM}_{2}$, which verifies that: $\nabla_{[a}X_{bc]}=0$ and
$\nabla_{b}X^{ab}=0$. The other class is: $\mathcal{EM}_{4} = \mathcal{EM}_{0}
- \mathcal{EM}_{1} \cup \mathcal{EM}_{2}$, which verifies that:
$\nabla_{[a}X_{bc]}\neq 0$ and $\nabla_{b}X^{ab}\neq 0$.

\begin{proposition}\label{p1}
If $\mathbf{X} \in \mathcal{EM}_{0}$, the third rank tensor $3
\nabla_{[a}X_{bc]}$ and the vector $(-1)\nabla_{b}\,^{*}X^{ab}$ are both dual.
Reciprocally, the tensor $3\nabla_{[a}\,^{*}X_{bc]}$ and the vector
$\nabla_{b}X^{ab}$ are also dual.
\end{proposition}

\begin{proposition}
If $\mathbf{X} \in \mathcal{EM}_{1}$, then $^{*}\mathbf{X} \in
\mathcal{EM}_{2}$ and reciprocally.
\end{proposition}

\begin{proof}
If $\mathbf{X} \in \mathcal{EM}_{1}$ then $d\mathbf{X}=0$,
therefore it is verified that: $\delta ^{*}\mathbf{X}=
^{*}d^{**}\mathbf{X}= - ^{*}d\mathbf{X}=0$. The reciprocal is also
right, if $\mathbf{X} \in \mathcal{EM}_{2}$ then
$\delta\mathbf{X}=0$, therefore it is verified that: $^{**}d
^{*}\mathbf{X}= (-1)^{2}d^{*}\mathbf{X}=0$.
\end{proof}

\begin{definition}
If $\mathbf{X} \in \mathcal{EM}_{0}$, and the vector $\mathbf{Z}$ is defined
as: $\mathbf{Z}=\delta \mathbf{X}$, that is: $Z_a = \nabla^{b}X_{ab}$, it is
called the $\alpha$-current of $\mathbf{X}$, and it verifies that: $\delta
\mathbf{Z} = 0 $.
\end{definition}

\begin{definition}
If $\mathbf{X} \in \mathcal{EM}_{0}$, and the vector $\mathbf{Y}$ is defined
as: $\mathbf{Y}=\delta ^{*}\mathbf{X}$, that is: $Y_a =
\nabla^{b}\,^{*}X_{ab}$, it is called the $\beta$-current of $\mathbf{X}$, and
it verifies that: $\delta \mathbf{Y} = 0 $.
\end{definition}

\begin{proposition}\label{p6}
If $\{\mathbf{ Z}, \mathbf{ Y} \}$ are the $\alpha$ and $\beta$-current of
$\mathbf{ X}$, then $\{\mathbf{ Y}, -\mathbf{ Z}\}$ are the $\alpha$ and
$\beta$-current of $\;^{*}\mathbf{ X}$.
\end{proposition}

\begin{theorem}[Maxwell-like]\label{maxwell}
Any tensor $\mathbf{X} \in \mathcal{EM}_{0}$ verifies the next equations, where
$\mathbf{Z}$ and $\mathbf{Y}$ are its $\alpha$ and $\beta$-current
respectively.
\begin{equation}
\delta \mathbf{X} = \mathbf{Z} \qquad \delta ^{*}\mathbf{X} = \mathbf{Y}
\end{equation}

verifying that $\delta \mathbf{Z}=0$ and $\delta \mathbf{Y}=0$.
\end{theorem}

%\begin{remark}
From Proposition \ref{p1} these equations can be expressed in tensorial form
as:
\begin{equation}
\nabla_{b} X^{ab} = Z^{a} \qquad \nabla_{[b} X_{cd]} =
-\frac{1}{3}\epsilon_{abcd}Y^a
\end{equation}

where $\nabla_{a}Z^{a}=0$ and $\nabla_{a}Y^{a}=0$. For the dual
$^{*}\mathbf{X}$, it is verified that:
\begin{equation}
\nabla_{b}\, ^{*}X^{ab} = Y^{a} \qquad \nabla_{[b}\,
^{*}X_{cd]}=\frac{1}{3}\epsilon_{abcd}Z^a
\end{equation}

these become the symmetrical Maxwell-like equations system and the current
conservation laws which are the starting point for the construction of the
electromagnetic duality. The $\mathcal{EM}_{1}$ or the $\mathcal{EM}_{2}$
classes are obtained if $\mathbf{Y}=0$ or $\mathbf{Z}=0$.
%\end{remark}

\begin{proposition}\label{potential}
If $\mathbf{X}\in \mathcal{EM}_{1}$, it can be expressed from a vector field
$\mathbf{U}$ as:
\begin{equation}
\mathbf{X}=d\mathbf{U} \Leftrightarrow X_{ab}= \nabla_{a} U_{b} - \nabla_{b}
U_{a}
\end{equation}
\end{proposition}

\begin{proof}
It follows from the Poincar\'e Lemma\cite{Gockeler:1989}(p 21),
\cite{Lovelock:1989}(p 141). In this case $\mathbf{X}$ is closed,
$d\mathbf{X}=0 $, then it must be exact, $\mathbf{X}=d\mathbf{U}$. However, the
vector field $\mathbf{U}$ is not univocally defined, being possible a
transformation as: $\mathbf{U}'=\mathbf{U}+d\phi$, where $\phi$ is a scalar
field. It is verified that: $d\mathbf{U}'=d\mathbf{U}$, due to: $dd\phi=0$. To
fix this scalar it is imposed an additional equation as the Lorentz gauge:
$\delta \mathbf{U}=0$. This restriction becomes:
\begin{equation}
\delta d \phi = 0 \Leftrightarrow -\square \phi =0
\end{equation}
\end{proof}

\begin{proposition}\label{existence}
If $\mathbf{Z}$ is a vector field verifying $\delta \mathbf{Z}=0$, then exist
solution for $\mathbf{X} \in \mathcal{EM}_{1}$ expressed as:
\begin{equation}
\delta \mathbf{X} = \mathbf{ Z} \qquad d\mathbf{X}=0
\end{equation}
\end{proposition}

\begin{proof}
From the double duality it is concluded that if a p-form is null also is null
its dual, so from $\delta \mathbf{Z}=\, ^{*}d^{*}\mathbf{Z}=0$ it is concluded
that ${*}\mathbf{Z}$ is closed. Based on the Poincar\'e Lemma it must be exact,
this is: $d ^{*}\mathbf{X} = \, ^{*}\mathbf{Z}$, therefore exist $\mathbf{X}$.
This is not univocally defined, it is verified that:
$\mathbf{X}'=\mathbf{X}+\mathbf{V}$, where $\mathbf{V} \in \mathcal{EM}_{3}$.
\end{proof}

\begin{theorem}[Split]\label{splitting}
Any field $\mathbf{X} \in \mathcal{EM}_{0}$ can be expressed by means of two
auxiliary $\alpha$-fields, $_{\alpha}X_{ab}$ and $_{\beta}X_{ab}$, such as:
\begin{eqnarray}
 X_{ab} & = & \, _{\alpha}X_{ab} - \;^{*}_{\beta}X_{ab} \\
 ^{*}X_{ab} & = & \, _{\beta}X_{ab} + \;^{*}_{\alpha}X_{ab}
\end{eqnarray}
where each auxiliary field verifies the following expressions based on the
$\alpha$ and $\beta$ currents of $\mathbf{X}$:
\begin{eqnarray}
\mathbf{\nabla} \cdot \; _{\alpha}\mathbf{X}=\mathbf{Z} & \qquad &
\mathbf{\nabla} \cdot \; _{\alpha}^{*}\mathbf{X}= 0 \label{sp1}\\
\mathbf{\nabla} \cdot \; _{\beta}\mathbf{X}=\mathbf{Y} & \qquad &
\mathbf{\nabla} \cdot \; _{\beta}^{*}\mathbf{X}= 0 \label{sp2}
\end{eqnarray}
\end{theorem}

\begin{proof}
Firstly if $\mathbf{X} \in \mathcal{EM}_{1}$ or $\mathbf{X} \in
\mathcal{EM}_{2}$ the theorem is immediately proved by choosing null any of the
two auxiliary fields. The other case is $\mathbf{X} \in \mathcal{EM}_{4}$ with
non-null sources. The auxiliary fields $_{\alpha}\mathbf{X}$ and
$_{\beta}\mathbf{X}$ are chosen according with the equations (\ref{sp1}) and
(\ref{sp2}). From the previous Proposition it is proved that there are solution
for both fields, however these solutions are not univocally defined. Therefore
these solutions can verify that:
\begin{equation}
D_{ab}= X_{ab} - \, _{\alpha}X_{ab} + \; ^{*}_{\beta}X_{ab}\neq 0.
\end{equation}

however it is obtained that $\mathbf{D}\in \mathcal{EM}_{3}$. Due to
$\mathcal{EM}_{3} \subset \mathcal{EM}_{1}$, it can be introduced a new tensor:
$_{\alpha}X'_{ab}=(_{\alpha}X_{ab}+D_{ab}) \in \mathcal{EM}_{1}$, consequently
verifying that: $X_{ab} = \, _{\alpha}X'_{ab} - \; ^{*}_{\beta}X_{ab}$.
\end{proof}

\begin{proposition}\label{noreduc}
If $\mathbf{A}$ and $\mathbf{B}$ belong to $\mathcal{EM}_{1}$ and not to
$\mathcal{EM}_{3}$, the solution for the equation:
\begin{equation}
\lambda\, ^{*}\mathbf{B}+ \sigma\mathbf{A}=0
\end{equation}
where $\lambda$ and $\sigma$ are constant, is that: $\lambda=\sigma=0$, i.e.
$\mathbf{A}$ and $^{*}\mathbf{B}$ are lineally independent.
\end{proposition}

\begin{proof}
From the previous conditions must be: $d\mathbf{A}=d\mathbf{B}=0$ and also
$\delta\mathbf{A} \neq 0$ and $\delta\mathbf{B}\neq 0$. However it is verified
that from: $\lambda d^{*}\mathbf{B} + \sigma d\mathbf{A}=0$, it must be:
$\lambda=0$. Also from: $\lambda \delta^{*}\mathbf{B} + \sigma
\delta\mathbf{A}=0$, and due that $\delta^{*}\mathbf{B}=-\,^{*}d\mathbf{B}=0$,
it must be: $\sigma=0$. Therefore in the general case the $\mathcal{EM}_{2}$
class is non-reducible to $\mathcal{EM}_{1}$ and reciprocally because both are
lineally independent.
\end{proof}

\begin{proposition}\label{copotential}
If $\mathbf{X} \in \mathcal{EM}_{3}$, it can be expressed by two co-potential
vectors, $\{U_{a}, V_{a}\}$ as:
\begin{equation}
X_{ab}=\nabla_{a}U_{b} - \nabla_{b}U_{a} \qquad ^{*}X_{ab}=\nabla_{a}V_{b} -
\nabla_{b}V_{a}
\end{equation}
\end{proposition}

\subsection{Bilinear Expressions}

Some useful properties of the tensor class $\mathcal{EM}_{0}$ can be obtained
by using bilinear expressions. All are based on a basic property related with a
partial double duality transformation.

\begin{proposition}\label{p10}
If the fourth rank tensor $\mathbf{ H}$ verifies that:
\begin{equation}
H_{abcd}=H_{[ab]cd}=H_{ab[cd]}
\end{equation}

It is verified that:
\begin{equation}
(^{*}H^{*})^{b}_{a} = \frac{1}{2} \epsilon_{acde} H^{de}_{\;\;\;uv} \frac{1}{2}
\epsilon^{bcuv} = H^{b}_{\;a} - \frac{1}{2}\delta^{b}_{a}H
\end{equation}

where: $H^{b}_{\;a} = H^{bc}_{\;\;\;ac}$ and $H = H^{cd}_{\;\;\;cd}$.
\end{proposition}
\begin{proof}
It is based on the properties of the $\mathbf{ \epsilon}$ and $\mathbf{
\delta}$ tensors\cite{Gockeler:1989}(p 75).
\end{proof}

\begin{proposition}\label{dual-exp}
If $\mathbf{A}$ and $\mathbf{B} \in \mathcal{EM}_{0}$, it is
verified that:
\begin{eqnarray}
 ^{*}A_{ac}\, ^{*}B^{bc} & = &  A_{ac}B^{bc} - \frac{1}{2}\delta^{b}_{a}
 A_{cd}B^{cd}\\
 A_{ac}\, ^{*}B^{bc} & = & - \, ^{*}A_{ac}B^{bc} + \frac{1}{2}\delta^{b}_{a} \,
 ^{*}A_{cd}B^{cd}\\
 ^{*}A_{cd}\, ^{*}B^{cd} & = & - A_{cd}B^{cd}\\
 A_{cd}\, ^{*}B^{cd} & = &  \, ^{*}A_{cd}B^{cd}
\end{eqnarray}
\end{proposition}

\begin{proof}
The first is obtained directly from the previous Proposition, the second is
obtained by using the intermediate step: $ A_{ac}\, ^{*}B^{bc} = - \,
^{**}A_{ac}\, ^{*}B^{bc} $. The two last expressions are contraction of the
previous.
\end{proof}

\begin{proposition}
If $\mathbf{ A}$ and $\mathbf{B} \in \mathcal{EM}_{0}$, it is verified that:
\begin{equation}
\nabla_{b}(A_{ac}B^{bc}) = A_{ca}(\nabla_{b} B^{cb}) +\, ^{*}B_{ca}(\nabla_{b}
\, ^{*} A^{cb}) +
 \frac{1}{2}B^{cd}\nabla_{a}(A_{cd})
\end{equation}
\end{proposition}

\begin{proof}
It is deduced directly applying the Theorem \ref{maxwell}.
\end{proof}

\begin{proposition}\label{p11}
Any $\mathbf{X} \in \mathcal{EM}_{0}$, its dual and currents verify that:
\begin{eqnarray}
 \nabla_{b}(X_{ac}X^{bc}) & = & Z^{c}X_{ca} + Y^{c}\;^{*}X_{ca} +
 \frac{1}{4}\nabla_{a}(X_{cd}X^{cd})\\
 \nabla_{b}(^{*}X_{ac}\,^{*}X^{bc}) & = &
 Y^{c}\;^{*}X_{ca} + Z^{c}X_{ca} + \frac{1}{4}\nabla_{a}(^{*}X_{cd}\,^{*}X^{cd})
\end{eqnarray}
\end{proposition}

\begin{theorem}[Stress-energy-like]\label{stress-energy}
If $\mathbf{X} \in \mathcal{EM}_{0}$, the three following expression are
equivalents:
\begin{eqnarray}
W^{b}_{a} & = & X_{ac}X^{bc} - \frac{1}{4}\delta^{b}_{a} X_{cd}X^{cd}\\
 & = & ^{*}X_{ac}\, ^{*}X^{bc}  -  \frac{1}{4}\delta^{b}_{a} \,^{*}X_{cd}\, ^{*}X^{cd}\\
 & = & \frac{1}{2}(X_{ac}X^{bc}  +   \, ^{*}X_{ac}\, ^{*}X^{bc})
\end{eqnarray}

And the symmetric second rank tensor $\mathbf{W}$ verifies that:
\begin{equation}
\nabla^{b}W_{ab}= Z^{b}X_{ba} + Y^{b}\, ^{*}X_{ba}
\end{equation}
\end{theorem}

\begin{proposition}\label{split}
The tensor $\mathbf{W}$ can be decomposed in three terms:
\begin{equation}
\mathbf{W}= \, _{(\alpha)}\mathbf{W} + \, _{(\beta)}\mathbf{W} + \,
_{(\alpha\beta)}\mathbf{W}
\end{equation}
where it is verified that:
\begin{eqnarray}
\nabla^{b}\, _{(\alpha)}W_{ab} & = & Z^{b}\, _{\alpha}X_{ba} \\ \nabla^{b}\,
_{(\beta)}W_{ab} & = & Y^{b}\, _{\beta}X_{ba} \\ \nabla^{b}\,
_{(\alpha\beta)}W_{ab} & = & - Z^{b}\, _{\beta}^{*}X_{ba} + Y^{b}\,
_{\alpha}^{*}X_{ba}
\end{eqnarray}
\end{proposition}

\begin{proof}
This result is based on the Theorem \ref{splitting}, each tensor is expressed
as:
\begin{eqnarray}
_{(\alpha)}W^{b}_{a} & = & \frac{1}{2}(_{\alpha}X_{ac}\, _{\alpha}X^{bc} + \,
_{\alpha}^{*}X_{ac}\, _{\alpha}^{*}X^{bc})\\ _{(\beta)}W^{b}_{a} & = &
\frac{1}{2}(_{\beta}X_{ac}\, _{\beta}X^{bc} + \, _{\beta}^{*}X_{ac}\,
_{\beta}^{*}X^{bc})\\ _{(\alpha\beta)}W^{b}_{a} & = &
\frac{1}{2}(_{\beta}X_{ac}\, _{\alpha}^{*}X^{bc} +\, _{\alpha}^{*}X_{ac}\,
_{\beta}X^{bc} - \, _{\alpha}X_{ac}\, _{\beta}^{*}X^{bc} - \,
_{\beta}^{*}X_{ac}\, _{\alpha}X^{bc})
\end{eqnarray}

The last tensor can be simplified by using: $N^{b}_{a} =
\;_{\beta}X_{ac}\, _{\alpha}^{*}X^{bc} - \, _{\alpha}X_{ac}\,
_{\beta}^{*}X^{bc}$. Based on Proposition \ref{dual-exp}, it is
verified that: $N_{ab}=N_{ba}$ and ${N^{a}}_{a}=0$. It is obtained
that: $ _{(\alpha\beta)}W_{ab} = (1/2)(N_{ab}+N_{ba}) = N_{ab}$.
\end{proof}

\section{The axiomatic equations of the electromagnetic field}

The class $\mathcal{EM}_{0}$ can be considered as an abstract and formal
representation of a dual electromagnetic field. Due to the existence of
magnetic monopoles is an hypothesis, any theoretical approach to this concept
must be based on some axiomatic. The only condition imposed in this paper is
that the field can be represented by a skew-symmetrical second rank tensor.

The electromagnetic field can be formally represented by a second
rank skew-symmetric tensor $\mathbf{ F} \in \mathcal{EM}_{0}$.
Based on Theorems \ref{maxwell}, \ref{splitting} and
\ref{stress-energy}, this tensor has some associated tensors: an
$\alpha$-current, $_{\alpha}\mathbf{ J}$, a $\beta$-current,
$_{\beta}\mathbf{ J}$, and a stress-energy tensor $\mathbf{ T}$.
They verify that:
\begin{eqnarray}
\mathbf{\nabla \cdot F} =4\pi\, _{\alpha}\mathbf{J} & \qquad & \mathbf{\nabla
\cdot}\, _{\alpha}\mathbf{ J} = 0 \label{cur1}\\ \mathbf{\nabla \cdot} \,
^{*}\mathbf{F} =4\pi\, _{\beta}\mathbf{J} & \qquad & \mathbf{\nabla \cdot}\,
_{\beta}\mathbf{ J} = 0 \label{cur2}\\
 T^{b}_{a}= \frac{1}{8\pi}(F_{ac}F^{bc} + \, ^{*}F_{ac}\, ^{*}F^{bc}) & \qquad &
  \mathbf{\nabla \cdot T} = \, _{\alpha}\mathbf{J \cdot F} +
  \, _{\beta}\mathbf{J \cdot} \, ^{*} \mathbf{F}
\end{eqnarray}

These equations are duality invariant, specifically the expression
of the stress-energy tensor match with a general invariant
form\cite{Deser:1996xp}. One of both the $\alpha$-current or the
$\beta$-current is an electric current and the other one is a
monopolar magnetic current. Both choice are valid because provide
the same formal symmetry to Maxwell equations. In this paper,
$_{\alpha}\mathbf{ J}$ and $_{\beta}\mathbf{ J}$ are considered
respectively as the electric and the magnetic monopolar currents.

A difference appears when this formal derivation is compared with
the classical theory of the electromagnetic field. Instead of a
potential vector, $\mathbf{ A}$, two potential vectors,
$_{\alpha}\mathbf{ A}$ and $_{\beta}\mathbf{ A}$, can be
considered based on the Theorem \ref{splitting}. Also two
auxiliary fields $ _{\alpha}\mathbf{ F}$ and $\,_{\beta}\mathbf{
F} \in \mathcal{EM}_{1}$ are introduced. They verify that:
\begin{eqnarray}
 \mathbf{ F} & = & _{\alpha}\mathbf{ F} - \,_{\beta}^{*}\mathbf{ F} \label{fi}\\
 _{\alpha} F_{ab} & = & \nabla_{a}\, _{\alpha}A_{b} - \nabla_{b}\,
 _{\alpha}A_{a}\label{fal}\\
 _{\beta} F_{ab} & = & \nabla_{a}\, _{\beta}A_{b} - \nabla_{b}\,
 _{\beta}A_{a}\label{fbe}
\end{eqnarray}

Each potential vector is independently related with its current
type as:
\begin{eqnarray}
  -  \square \, _{\alpha}A_{a}  +  R^{b}_{\;\;a}\, _{\alpha}A_{b} & =  & 4\pi \, _{\alpha}J_{a} \label{aal}\\
  -  \square \, _{\beta} A_{a}  +   R^{b}_{\;\;a}\, _{\beta} A_{b} & = & 4\pi \, _{\beta}
  J_{a}\label{abe}
\end{eqnarray}

In an electromagnetic duality hypothesis, two independent currents
must be considered: the magnetic and the electric sources. Two
approaches can be considered for mapping these sources in an
electromagnetic field. In this paper, similarly to other works on
electromagnetic duality\cite{Zwanziger:1971hk, Mukherjee:1998he},
two vector fields are considered, this choice preserves the radial
vs axial properties of fields. The magnetic field generated by a
magnetic charge is of radial type based on the gradient of a
scalar, while the magnetic field generated by a electric charge is
of axial type based on a curl. In a full electromagnetic duality
two radial and two axial field are involved. Most works in
electromagnetic duality fit the dual sources into a single vector
field. This task needs of some heuristic as the Dirac string
monopole, which is a topological construction\cite{Wu:1975vq,
Thober:1999xb}.

The existence of magnetic monopoles is an hypothesis that extended
the symmetry of Maxwell equations towards the symmetry of
charges\cite{Weinberg:1998aw}. The field generated by a singular
electric charge is well mapped into a vector potential. While
that, if the field generates by a magnetic monopole is mapped in
the same vector potential, the expected symmetry is broken. It
appears two different concepts: the point singular electric charge
and the string singular magnetic monopole. The Dirac quantization
arises as a condition imposed to avoid the physical effect of the
string\cite{Weinberg:1998aw} \cite{Gockeler:1989}(p 165).

If a full symmetrical duality is desired, the topological
counterpart of a point singular electric charge must be other one.
Based on the Theorem \ref{splitting} a simple and symmetrical
solution is found by using two potential vectors. This approach
has some advantages because a switch between the fields $\mathbf{
F} \rightarrow\, ^{*}\mathbf{ F}$ is reduced, as shown equation
(\ref{fi}), to a switch between the two auxiliary fields:
$_{\alpha}\mathbf{ F} \rightarrow\, _{\beta}\mathbf{ F}$ and
$_{\beta}\mathbf{ F} \rightarrow\,- \; _{\alpha}\mathbf{ F}$,
which is more similar to the current switch: $_{\alpha}\mathbf{
J}\rightarrow\, _{\beta}\mathbf{ J}$ and $_{\beta}\mathbf{
J}\rightarrow\,- \; _{\alpha}\mathbf{ J}$ as is deduced from
equations (\ref{cur1}) and (\ref{cur2}).

\section{Duality Invariance}

The duality invariance is related to how field properties are
modified with the transposition: $\mathbf{ F} \rightarrow\,
^{*}\mathbf{ F}$. Some equation changes are expected with this
transposition. Invariant equations are preferred because it is
supposed that both representations, $\mathbf{ F}$ and
$^{*}\mathbf{ F}$ are equivalents. The main principle involved in
duality invariance is that both descriptions have the same
information, therefore both descriptions must be valid
representations of the physical fact. Non privileged reference
system can be found in General Relativity to study the physical
laws, consequently all reference system are equivalent for this
propose. By analogy, any symbol which have the same information
should be equivalent to study the physical laws.

Instead of invariance in a discrete transposition as $\mathbf{ F}
\rightarrow\, ^{*}\mathbf{ F}$, it is required a continuous
invariance related with a rotation phase $\phi$ between the two
fields. A complex tensor can be introduced to deal with this
rotations: $\mathbf{ F} + \imath ^{*}\mathbf{ F}$. In this case a
rotation generates the complex tensor: $ e^{\imath \phi}(\mathbf{
F} + \imath ^{*}\mathbf{ F})$. Duality invariance must be
considered as mathematical invariance to rotations. As well as in
General Relativity the covariant representation is introduced to
be independent of any coordinate systems, a representation
invariant to phase $\phi$ must be introduced. Equivalent to the
representation in the complex plane, a continuous transformation
is introduced as:
\begin{equation}
\mathbf{ F}'=\mathbf{ F} \cos \phi + ^{*}\mathbf{ F} \sin\phi
\end{equation}

By taking its dual: $^{*}\mathbf{ F}'=\,^{*}\mathbf{ F} \cos \phi
- \mathbf{ F} \sin\phi $, it is obtained a representation based on
a rotation matrix, $\mathbf{ R} (\phi)$, similar to the used by
Deser et al.\cite{Deser:1997xu, Deser:1997mz}:
\begin{equation}
\left (
\begin{array}{r}
  \mathbf{ F} \\
 ^{*}\mathbf{ F}
\end{array}
\right ) ' = \mathbf{ R} (\phi) \left (
\begin{array}{r}
  \mathbf{ F} \\
 ^{*}\mathbf{ F}
\end{array}
\right ) \qquad \mathbf{ R}(\phi) = \left (
\begin{array}{cc}
  \cos\phi & \sin\phi \\
  -\sin\phi & \cos\phi
\end{array}
\right )
\end{equation}

A duality plane, $D=2$, is introduced in addition to the Riemannian space-time
with $D=4$. This approach clearly evidences the well known connection between
the duality problem and the rotation invariance, in this case with the SO(2)
group. The tensors having two parts, e.g. $\alpha$ and $\beta$, can be
represented by extended tensors with uppercase index:
\begin{equation}
L^{Ai...}_{a...}\quad M^{Ai...}_{a...}\qquad A=0,1\quad a,...,i,...=0,1,2,3
\end{equation}

Any extended tensor, $L^{Aa}_{b}$, can be expressed symbolically
in short as: $\mathbf{L}^{A}$, $\mathbf{L}^{a}_{b}$ and also as
$\mathbf{L}$, by abstracting some or all the index types. The
covariant and contravariant uppercase index are related with the
symmetrical group while the lowercase index are related with the
space-time. E.g. the electromagnetic field tensors, which have a
dual interpretation, are represented by the extended tensors:
$F^{C}_{ab}$, $A^{C}_{a}$ and $J^{C}_{a}$, such as: $ {
F}^{0}_{ab} = { F}_{ab}$ and ${ F}^{1}_{ab} = \, ^{*}{ F}_{ab}$.
The Maxwell equations are:
\begin{equation}
\delta \left (
\begin{array}{r}
  \mathbf{ F} \\
 ^{*}\mathbf{ F}
\end{array}
\right ) = 4\pi \left (
\begin{array}{r}
  _{\alpha}\mathbf{ J} \\
 _{\beta}\mathbf{ J}
\end{array}
\right )
\end{equation}

It is obtained that:
\begin{equation}
\left (
\begin{array}{r}
  _{\alpha}\mathbf{J} \\
  _{\beta}\mathbf{J}
\end{array}
\right ) ' = \mathbf{ R} (\phi) \left (
\begin{array}{r}
  _{\alpha}\mathbf{J} \\
  _{\beta}\mathbf{J}
\end{array}
\right )
\end{equation}

The concept of rotation in the $\mathcal{EM}_{0}$ class is extended to all
$\alpha$ and $\beta$ related tensors. They are transformed as follows:
\begin{equation}
\mathbf{ L} ^{'i...}_{a...} = \mathbf{ R}(\phi) \mathbf{ L}^{i...}_{a...}
\end{equation}

In the duality plane the symmetrical identity tensor, $\eta_{AB}$,
and the skew-symmetrical tensor, $\epsilon_{AB}$ are considered.
They are represented as:
\begin{equation}
\mathbf{ \eta} = \left (
\begin{array}{cc}
  1 & 0 \\
  0 & 1
\end{array} \right ) \qquad
\mathbf{ \epsilon} = \left (
\begin{array}{rr}
  0 & 1 \\
  -1 & 0
\end{array} \right )
\end{equation}

If $\mathbf{\epsilon}^{T}$ is the transpose of
$\mathbf{\epsilon}$, it is verified that:
$\mathbf{\epsilon}^{T}\mathbf{\epsilon}=\mathbf{\eta}$, and also:
$\mathbf{\epsilon}\mathbf{\epsilon}= - \mathbf{\eta}$. The metric
tensor allows the use of covariant and contravariant components
related to: $A^{i...}_{Ca...} = \eta_{CD} A^{Di...}_{a...}$. Two
operators are introduced in the duality plane. The following
expression are called the scalar and the vectorial products in the
duality plane, which are operators based on the previous matrices:
\begin{eqnarray}
\mathbf{ A}^{i...}_{a...} \Ddot \mathbf{ B}^{j...}_{b...} & = &
\eta_{CD}A^{Ci...}_{a...}B^{Dj...}_{b...} = < \mathbf{
A}^{i...}_{a...}, \mathbf{ B}^{j...}_{b...} > \\ \mathbf{
A}^{i...}_{a...} \Dwedge \mathbf{ B}^{j...}_{b...} & = &
\epsilon_{CD} A^{Ci...}_{a...} B^{Dj...}_{b...} = [ \mathbf{
A}^{i...}_{a...}, \mathbf{ B}^{j...}_{b...} ]
\end{eqnarray}

These bilinear expressions are duality invariant because if $\mathbf{ R}^{T}$
is the transpose of the rotation matrix, it is verified the following classical
expressions of the rotations groups:
\begin{equation}
\mathbf{ R}^{T} (\phi) \mathbf{ \eta} \mathbf{ R} (\phi) = \mathbf{ \eta}
\qquad \mathbf{ R}^{T} (\phi) \mathbf{ \epsilon} \mathbf{ R} (\phi)
 = \mathbf{ \epsilon}
\end{equation}

Any matrix, $\mathbf{ \lambda}$, which generates a bilinear
invariant operator must verify: $\mathbf{ R}^{T} (\phi) \mathbf{
\lambda} \mathbf{ R} (\phi) = \mathbf{ \lambda}$, but any solution
of this equation can be expressed as a linear combination of the
previous: $\mathbf{ \lambda} = a \mathbf{ \eta} + b \mathbf{
\epsilon}$, where $a$ and $b$ are constant. This is due to:
\begin{equation}
\mathbf{ R} (\phi) = \mathbf{ \eta}\cos\phi + \mathbf{\epsilon}\sin\phi \qquad
\mathbf{ R}^{T} (\phi) = \mathbf{\eta}\cos\phi - \mathbf{\epsilon}\sin\phi
\end{equation}

Therefore, any bilinear invariant in the duality plane can be
constructed from the defined scalar and vectorial operators.  The
scalar and vectorial products are symmetric and skew-symmetric
respectively. Both are linear operators, symbolically expressed
as:
\begin{eqnarray}
{\mathbf{ L}}\Ddot (c{\mathbf{ M}} + d{\mathbf{ N}} ) & = & c({\mathbf{ L}}
\Ddot {\mathbf{ M}}) + d({\mathbf{ L}} \Ddot {\mathbf{ N}})
\\ {\mathbf{ L}}\Dwedge (c{\mathbf{ M}} + d{\mathbf{ N}} ) & = & c ({\mathbf{
L}} \Dwedge {\mathbf{ M}}) + d({\mathbf{ L}} \Dwedge {\mathbf{ N}})
\end{eqnarray}

where $(c, d)$ are ordinary tensors, which are constant for the duality plane
operators. Extended tensors allows to express the electromagnetic equations by
using more compact and invariant expressions. The equations (\ref{aal}) and
(\ref{abe}) are expressed as:

\begin{equation}
-\square \, { A}^{C}_{a}+{R^{b}}_{a}\, { A}^{C}_{b}
 = 4\pi{ J}^{C}_{a}
\end{equation}

The $ {\mathbf{ K}}$ field is introduced to deal more easily with
the first derivative of the potential vector, it is defined as
follows:

\begin{equation}
K^{C}_{ab} = \nabla_{a}\, A^{C}_{b} - \nabla_{b}\,{ A}^{C}_{a}
\end{equation}

The equation (\ref{fi}) is represented as:

\begin{equation}\label{field1}
{ \mathbf{ F}}^{C} = { \mathbf{ K}}^{C} - \epsilon^{CD}\; ^{*}{\mathbf{ K}}_{D}
\end{equation}

Remark that the fields $\mathbf{F}^{0}$ and $\mathbf{F}^{1}$ are
dependent due to: $\mathbf{F}^{1}=\, ^{*}\mathbf{F}^{0}$, while
$\mathbf{K}^{0}$ and $\mathbf{K}^{1}$ are two independent fields.
It is possible to express the Maxwell equations and the current
conservation law in a compact and invariant form as:
\begin{equation}
\mathbf{\nabla} \cdot \mathbf{ F}^{C} =4\pi {\mathbf{ J}}^{C} \qquad
\mathbf{\nabla} \cdot {\mathbf{ J}}^{C} = 0
\end{equation}

Finally, the stress-energy tensor $ {\mathbf{ T}}$ can be expressed in duality
invariant form as:
\begin{equation}\label{force}
 T^{b}_{a}  =  \frac{1}{8\pi}({\mathbf{ F}}_{ac} \Ddot
{\mathbf{ F}}^{bc}) \qquad
 \mathbf{\nabla \cdot T}  = \mathbf{\mathcal{F}}=
 {\mathbf{ J}} \Ddot
{\mathbf{ F}}
\end{equation}

Where $\mathbf{\mathcal{ F}}$ is the force density produced by the
field-current interaction. When a duality plane rotation is
produced with $\phi = \pi/2$, a permutation between the electric
and magnetic field is verified in the field ${\mathbf{ F}}$, but
this rotation does a permutation between $\alpha$ and $\beta$
parts in the extended tensors: ${\mathbf{ J}}$, ${\mathbf{ A}}$
and ${\mathbf{ K}}$. This last belongs to $\mathcal{EM}_{1}$ and
verifies the following Maxwell equations:
\begin{equation}\label{soft}
\mathbf{\nabla \cdot} {\mathbf{ K}}^{C} = 4\pi{\mathbf{ J}}^{C} \qquad
\mathbf{\nabla \cdot} \;^{*}{\mathbf{ K}}^{C} = 0
\end{equation}

Based on Proposition \ref{split}, the stress-energy tensor can be expressed as:
$\mathbf{ T} = \, _{(\alpha)}\mathbf{ T} + \, _{(\beta)}\mathbf{ T} + \,
_{(\alpha\beta)}\mathbf{ T}$. The two first terms can be merged in a more
compact representation called the parallel stress-energy tensor: $\mathbf{
T}_{\parallel}$:
\begin{equation}\label{aplusb}
T^{b}_{\parallel a} = \, _{(\alpha)}T^{b}_{a} + \, _{(\beta)}T^{b}_{a} =
\frac{1}{8\pi}({\mathbf{ K}}_{ac} \Ddot {\mathbf{ K}}^{bc} + \;^{*}{\mathbf{
K}}_{ac} \Ddot \, ^{*}{\mathbf{ K}}^{bc})
\end{equation}

which verifies:
\begin{equation}
\mathbf{\nabla \cdot} {\mathbf{ T}}_{\parallel} =
\mathcal{F}_{\parallel}={\mathbf{ J}} \Ddot {\mathbf{ K}}
\end{equation}

where ${\mathbf{ J}} \Ddot {\mathbf{ K}}= \mathbf{ J}^{0} \cdot
\mathbf{ K}^{0} + \mathbf{ J}^{1} \cdot \mathbf{ K}^{1}$ is the
parallel force density. In this vase, the electrical current is
interacting with the electrical generated field and reciprocally
for the magnetic current. This case is composed of two independent
and separable electromagnetic fields. Each field is produced by
its own current, which is neutral to the action of the other
field. The term $_{(\alpha \beta)}\mathbf{ T}$ is called the
perpendicular stress-energy tensor. Based on the proposition
\ref{split}, it can be expressed as:

\begin{equation}\label{atimesb}
T^{b}_{\perp a} = - \frac{1}{8\pi}({\mathbf{ K}}_{ac} \Dwedge \, ^{*}{\mathbf{
K}}^{bc} + {\mathbf{ K}}^{bc} \Dwedge \, ^{*}{\mathbf{ K}}_{ac}) = -
\frac{1}{4\pi}{\mathbf{ K}}_{ac} \Dwedge \, ^{*}{\mathbf{ K}}^{bc}
\end{equation}

It verifies that:
\begin{equation}\label{perp}
 \mathbf{\nabla} \cdot {\mathbf{ T}}_{\perp} = \mathcal{F}_{\perp}= - {\mathbf{ J}} \Dwedge \, ^{*}{\mathbf{ K}}
\end{equation}

where ${\mathbf{ J}} \Dwedge \, ^{*}{\mathbf{ K}}= \mathbf{ J}^{0}
\cdot \, ^{*}\mathbf{ K}^{1} - \mathbf{ J}^{1} \cdot \,
^{*}\mathbf{ K}^{0}$ is the perpendicular force density. In this
case the electric current is interacting with the magnetic
generated field being neutral to the electrical generated field,
and reciprocally for the magnetic current.

\section{Lagrange function invariance}

A complete electromagnetic theory must include a model of the Lagrange action
function. An invariant action must be proposed to solve the variational problem
$\delta I =0$, where:
\begin{equation}
I=\int \sqrt{-g}L(\mathbf{ A}, \mathbf{ F}) d^{4}x
\end{equation}

In classical theory the Lagrange function $L$ is proportional to
$F_{ab}F^{ab}$, however this expression is not duality invariant.
Unfortunately all the following expressions are null:
\begin{equation}
{\mathbf{ F}}_{ab} \Ddot {\mathbf{ F}}^{ab} = {\mathbf{ F}}_{ab}
\Ddot {^{*}\mathbf{ F}}^{ab} = {^{*}\mathbf{ F}}_{ab} \Ddot
{^{*}\mathbf{ F}}^{ab} =0
\end{equation}
\begin{equation}
{\mathbf{ F}}_{ab} \Dwedge {\mathbf{ F}}^{ab} = {\mathbf{ F}}_{ab}
\Dwedge {^{*}\mathbf{ F}}^{ab} = {^{*}\mathbf{ F}}_{ab} \Dwedge
{^{*}\mathbf{ F}}^{ab} =0
\end{equation}

The Lagrange function must verify three imposed restrictions.
Firstly it must be duality invariant. The second restriction is
that the Maxwell equations must be obtained with variations of the
potential vector $\delta \mathbf{ A}$. Finally, the function must
provide the correct expression of the stress-energy tensor with
the metric variations $\delta g^{ab}$. A lot of works are recently
concerning with the construction of a Lagrange function verifying
these restrictions \cite{Wotzasek:1998rj, Bhattacharyya:1998fq,
Igarashi:1998hd, Deser:1997mz}.

In the approach of this paper two vector fields are considered,
therefore the function must be expressed as: $L(\mathbf{ A}^{C},
\mathbf{ K}^{C})$. The Lagrange function related with the field is
composed of two terms, the first is concerning with the
electromagnetic field itself, and the second with the
current-field interaction:
\begin{equation}
L = L_{F}({\mathbf{K}}^{C}) + L_{I}({\mathbf{ J}}^{C},{\mathbf{ A}}^{C})
\end{equation}

In order to generate the Maxwell equations, the interaction term
is based on the expression: ${\mathbf{ J}}_{a} \Ddot {\mathbf{
A}}^{a}$. Due to $\mathbf{T}=\mathbf{T}_{\parallel}+
\mathbf{T}_{\perp}$, the Lagrange function can be also expressed
as: $L_{F}=L_{\parallel}+L_{\perp}$. The field term can be
constructed by using bilinear invariant expressions formed with
the tensors ${\mathbf{ K}}$ and $^{*}{\mathbf{ K}}$. However the
following vectorial-based bilinear expressions are null.
\begin{equation}
{\mathbf{ K}}_{ab} \Dwedge {\mathbf{ K}}^{ab}={\mathbf{ K}}_{ab} \Dwedge\,
^{*}{\mathbf{ K}}^{ab} = \, ^{*}{\mathbf{ K}}_{ab} \Dwedge \, ^{*}{\mathbf{
K}}^{ab} = 0
\end{equation}

The scalar-based bilinear expressions are:
\begin{equation}
\sigma={\mathbf{ K}}_{ab} \Ddot {\mathbf{ K}}^{ab} \qquad
\tau={\mathbf{ K}}_{ab} \Ddot \, ^{*}{\mathbf{ K}}^{ab} \qquad
^{*}{\mathbf{ K}}_{ab} \Ddot \, ^{*}{\mathbf{ K}}^{ab} = - \,
{\mathbf{ K}}_{ab} \Ddot {\mathbf{ K}}^{ab}
\end{equation}

The $L_{F}$ term must be dependent of these expressions:
$L_{F}(\sigma,\tau)$. In a linear electromagnetic theory must be:
$\partial_{\mathbf{K}} L= c \mathbf{K}$, consequently an
expression as: $L_{F}=c_{\sigma}\sigma + c_{\tau}\tau$ must be
considered. However, $L_{\perp}$ can not be obtained with $\sigma$
and $\tau$ expressions, and also the $\tau$ expression can be
expressed as a gradient and removed from the Lagrange function:
\begin{equation}
\tau={\mathbf{ K}}_{ab} \Ddot \, ^{*}{\mathbf{ K}}^{ab}= 2
\nabla_{a}(\epsilon^{abcd}{\mathbf{ A}}_{b} \Ddot
\nabla_{c}{\mathbf{ A}}_{d})
\end{equation}

The term $\sigma={\mathbf{ K}}_{ab} \Ddot {\mathbf{ K}}^{ab}$ is
the only non-null, duality invariant and valid bilinear expression
that generates a linear electromagnetic theory. It generates the
Maxwell equations and provides the parallel stress-energy tensor.
The $L_{F}$ term can be expressed as:

\begin{equation}\label{gfactor}
L_{F} = L_{\parallel} + \gamma L_{\perp} \qquad \gamma \in \{0,1
\}
\end{equation}

where the $\gamma$ constant is related with the inclusion of a
perpendicular term in the Lagrangian. A solution for the the
Lagrange function including the interaction term $L_{I}$ is
expressed as:
\begin{equation}
L_{\parallel} + L_{I}= - \frac{1}{16\pi} {\mathbf{ K}}_{ab} \Ddot
{\mathbf{ K}}^{ab} + {\mathbf{ J}}_{a} \Ddot {\mathbf{ A}}^{a}
\end{equation}

The variation $\delta { A}^{Ca}$ generates the following
Euler-Lagrange conditions:
\begin{equation}
\frac{\partial { L}}{\partial { A}^{Ca}}= 2 \nabla^{b} \left [\frac {\partial
L}{\partial { K}^{Cba}} \right ]
\end{equation}

which provide the Maxwell equations: $4\pi{ J}^{C}_{a}= \nabla^{b} {
K}^{C}_{ab}$. Based on equations (\ref{field1}) and (\ref{soft}), it is
verified that:
\begin{equation}
\nabla^{b} { F}^{C}_{ab}= \nabla^{b} { K}^{C}_{ab} - \epsilon^{CD} \nabla^{b}
\;{ ^{*}K}_{Dab} = \nabla^{b} { K}^{C}_{ab} = 4\pi{ J}^{C}_{a}
\end{equation}

According to Einstein General Relativity, the stress-energy tensor can be
obtained from the field action as:
\begin{equation}
T_{ab} = -2 \frac{\partial L_{F} }{\partial g^{ab}} + g_{ab}L_{F}
\end{equation}

It is obtained the next result:
\begin{equation}
T^{b}_{\parallel a} = \frac{1}{4\pi}{\mathbf{ K}}_{ac} \Ddot {\mathbf{ K}}^{bc}
- \frac{1}{16\pi} \delta^{b}_{a} {\mathbf{ K}}_{cd} \Ddot {\mathbf{ K}}^{cd}
\end{equation}

Based on the Proposition \ref{dual-exp}, it can be expressed as in
the equation (\ref{aplusb}). By using the Einstein equation,
$\mathbf{ G} = 8\pi \mathbf{ T}$, this expression of the
stress-energy tensor allows to obtain the metric and field
associated to a charged central body as a black hole. In the outer
space of the body, the field can be expressed as: ${ K}^{C}_{ab}=
{ Q}^{C} K_{ab}$, where $ {\mathbf{ Q}}$ is the body charge, and
$K_{ab}$ is the solution for an unitary charge. The perpendicular
stress-energy tensor is proportional to the term: $\mathbf{Q}
\Dwedge \mathbf{Q}$ which is null. The stress-energy tensor is
only based on the scalar product of the charge, being multiplied
by $(Q)^{2}= {\mathbf{ Q}} \Ddot {\mathbf{ Q}}$, where $Q$ is the
equivalent combined charge. The Reissner-Nordstr\"{o}n
metric\cite{Hawking:1995ap, Misner:1973} arises as the solution
for this problem based on the mass $M$ and the combined charge:
$(Q)^{2}$. It is concluded that from the viewpoint of the
electromagnetic duality presented in this paper, the metric of an
electric or magnetic charged black hole are identical formally but
different numerically.

\section{Particle-Field Interaction}

From an operational viewpoint, a field is the formal
representation of a physical fact which determines an interaction
with some type of particle. A trajectory $x^{a}(s)$ is associated
to each virtual particle. If non-quantum theory is considered, the
trajectory is the formal representation of the virtual particle. A
vector field $\mathbf{ p}$ tangent to the trajectory is
considered. If the particle is massive, $p= \| \mathbf{ p} \| \neq
0$, the trajectory has an unitary tangent vector $\mathbf{ u}$.
The absolute derivation of $\mathbf{p}$ along a field line can be
expressed as follows:
\begin{equation}
\frac{d \mathbf{ p}}{ds}= \frac{D p^{a}}{ds}\mathbf{e}_a = (\nabla_{b} p^{a})
u^{b} \mathbf{e}_a
\end{equation}

Due to $p_{a}p^{a}$ is constant, it can be obtained the next expression:
$(\nabla_{b} p_{a})u^{b}=2\nabla_{[b} p_{a]}u^{b}$. It is concluded that the
vector field $\mathbf{ p}$ with non-null and constant $p= \| \mathbf{p} \| $
verifies the next equation along any of its field lines:
\begin{equation}\label{lorentz}
\frac{Dp^{a}}{ds} = f^{a} = p^{a}_{\;\;b}u^{b}
\end{equation}

Where $ p_{ab} \in \mathcal{EM}_{1}$ can be expressed as:
$p_{ab}=2 \nabla_{[b} p_{a]}$. The vector $\mathbf{ f}$ is the
Lorentz force based on a coupling constant or particle charge,
which is represented by a constant vector in the duality
plane\cite{Olive:1995sw}: ${\mathbf{ q}}\equiv( _{\alpha}q, \,
_{\beta}q)$. It is transformed by a rotation
as\cite{Deser:1996xp}: $\mathbf{ q}' = \mathbf{ R} (\phi) \mathbf{
q}$. When a phase rotation with $\phi=\pi/2$ is performed, the
charge is changed from $(_{\alpha}q,\, _{\beta}q)$ to
$(_{\beta}q,- \, _{\alpha}q)$. The term $\mathcal{F}={\mathbf{ J}}
\Ddot {\mathbf{ F}}$ is the force density in a continuous case. It
can be expressed as: $\mathcal{F}= \mathcal{F}_{\parallel} +
\gamma \mathcal{F}_{\perp}$, its expression is:
\begin{equation}\label{f2}
\mathcal{ F}_{a} = ({\mathbf{ J}}^{b}) \Ddot {\mathbf{ K}}_{ab} -
\gamma ( {\mathbf{ J}}^{b}) \Dwedge \, ^{*}{\mathbf{ K}}_{ab}
\end{equation}

By analogy, for a virtual particle, the term
$\mathbf{j}^{a}={\mathbf{ q}} u^{a}$ can be considered as the
discrete particle current, and the expression:
$\mathbf{f}=({\mathbf{ q}} \Ddot {\mathbf{ F}})\cdot \mathbf{ u}$
as the force of the field-particle interaction, its expression is:

\begin{equation}\label{f1}
f_{a} = ({\mathbf{ q}} u^{b}) \Ddot {\mathbf{ K}}_{ab} - \gamma (
{\mathbf{ q}} u^{b}) \Dwedge \, ^{*}{\mathbf{ K}}_{ab}
\end{equation}

which also is composed of the parallel and the perpendicular
force. From equations (\ref{lorentz}) and (\ref{f1}), the next
equation must be considered to solve the particle trajectory:

\begin{equation}\label{heavy}
p_{ab} = {\mathbf{ q}} \Ddot {\mathbf{ K}}_{ab} - \gamma {\mathbf{
q}} \Dwedge \, ^{*}{\mathbf{ K}}_{ab}
\end{equation}

which can be transformed as:

\begin{equation}\label{condition}
\nabla_{[b}(\mathbf{ p} +{\mathbf{ q}} \Ddot {\mathbf{ A}})_{a]}=
- \gamma \, ^{*}({\mathbf{ q}} \Dwedge {\mathbf{ K}}_{ab})
\end{equation}

This equation is non-homogeneous because the lhs is in
$\mathcal{EM}_{1}$ and the rhs is in $\mathcal{EM}_{2}$. Based on
the Proposition \ref{noreduc}, there are not a general solution
for this equation. A particular solution is possible if $\gamma
{\mathbf{ q}} \Dwedge {\mathbf{ K}}_{ab} \in \mathcal{EM}_{3}$,
but this implies that: $\gamma {\mathbf{ q}} \Dwedge {\mathbf{
J}}=0$.

A particular case can be obtained by considering the interaction
of two particles. One is acting as the field source, its field
tensors are proportional to the charge ${\mathbf{ q}}'$. The other
particle is acting as a singular test particle with charge
$\mathbf{ q}$. If the electric and magnetic charges are supposed
basic values multiplied by integer numbers, being $e$ and $\mu$
the basic electric and magnetic values respectively. For a
particle pair, the scalar and the vectorial product of the charges
are two interaction constants similar to the charge product in
classical electrodynamic:

\begin{equation}\label{quan}
{\mathbf{ q}} \Dwedge {\mathbf{ q}}' = (e\mu)n \qquad {\mathbf{ q}} \Ddot
{\mathbf{ q}}' = (e)^{2}m + (\mu)^{2}k \qquad n, m, k= 0, \pm1, \pm2,\ldots
\end{equation}

The vectorial product $ {\mathbf{ q}} \Dwedge {\mathbf{ q}}'=\,
^{0}q \, ^{1}q' - \, ^{1}q \, ^{0}q' $ is in the classical theory
the equivalent to the corresponding in the quantum theory known as
the Zwanziger-Schwinger charge quantization\cite{Lee:1997pn,
Olive:1995sw, Zwanziger:1968rs, Schwinger:1969ib} which is an
extension of the Dirac quantization. The field out of the source
particle can be expressed as: $\mathbf{K}_{ab}=\mathbf{q}'
K_{ab}$, with $K_{ab} \in \mathcal{EM}_{3}$. Based on Proposition
\ref{copotential} it is obtained: $K_{ab} = \nabla_{a}A_{b} -
\nabla_{b}A_{a}$ and also: $^{*}K_{ab}=
\nabla_{a}U_{b}-\nabla_{b}U_{a}$. The equation (\ref{condition})
becomes:

\begin{equation}\label{condition2}
\nabla_{[b}(\mathbf{ p} +{\mathbf{ q}} \Ddot {\mathbf{
q}'}\mathbf{A} + \gamma {\mathbf{ q}} \Dwedge {\mathbf{
q}'}\mathbf{U})_{a]}= 0
\end{equation}

This equation can be immediately solved by means of a scalar gradient as:
\begin{equation}
{ p}_{a} ={ \nabla}_{a} S - ({\mathbf{ q}} \Ddot {\mathbf{
q'})A_{a}} - \gamma ({\mathbf{ q}} \Dwedge {\mathbf{ q'})U_{a}}
\end{equation}

Where $S$ is the particle action. Consequently, the Hamilton-Jacobi approach is
obtained for the dynamic of the test particle, based on two co-potential
vector, $A_{a}$ and $U_{a}$. By using the two constants: $\theta_{\parallel}=
\mathbf{ q} \Ddot \mathbf{ q}'$ and $\theta_{\perp}= \mathbf{ q} \Dwedge
\mathbf{ q}'$, it is obtained:

\begin{equation}
(\nabla_{a} S - \theta_{\parallel} A_{a} - \gamma \theta_{\perp}
U_{a})(\nabla^{a} S - \theta_{\parallel} A^{a} - \gamma
\theta_{\perp} U^{a}) = (p)^{2}
\end{equation}

A solution with $\gamma=0$ is valid, but in this case the magnetic
monopole with charge: $(0,\pm\mu)$ is neutral to the electric
charge $(\pm e,0)$. In this case, due to equations (\ref{f2}) and
(\ref{f1}), the operational field responsible of the current-field
and particle-field interaction is the set of two independent
$\mathcal{EM}_{1}$ fields $\mathbf{K}$, instead of the
$\mathcal{EM}_{0}$ field $\mathbf{F}$.

\section{Conclusions}

A theory of electromagnetic duality has been presented. It has
been developed in the framework of the Einstein General
Relativity. The proposed theory carries out the symmetry of
Maxwell equations, the invariance of the stress-energy tensor, and
also the invariance of part of the Lagrange action. The theory has
been developed using an axiomatic method to deduce the dual
Maxwell equations from the properties of differential forms. Two
potential vectors are needed to deal with the two source types
involved in the electromagnetic duality.

The main result of this paper is that the stress-energy tensor can
be decomposed on two parts: the parallel and the perpendicular. In
the parallel part each current type is interacting with the field
of the same type. In the perpendicular part the interaction
becomes between currents and fields of different type. The
perpendicular part can not be easily integrated in a Lagrange
approach. On a linear electromagnetic theory, constructed with
duality invariant bilinear expressions, do not exist a suitable
expressions for this function. A Lagrange function with the
parallel part alone is a valid solution, it generates a
non-standard model of magnetic monopoles neutral to the electric
charges.

\end{document}